\documentclass[preprint2]{aastex}

\DeclareRobustCommand{\ion}[2]{\textup{#1\,\textsc{\lowercase{#2}}}}  

\begin{document}


\title{AGN luminosity and stellar age -- two missing ingredients for AGN unification as seen with iPTF supernovae}


\author{Beatriz Villarroel\altaffilmark{1,2}, Anders Nyholm\altaffilmark{3}, Torgny Karlsson\altaffilmark{1}, S\'ebastien Comer\'on\altaffilmark{4}, Andreas J. Korn\altaffilmark{1}, Jesper Sollerman\altaffilmark{3}, Erik Zackrisson\altaffilmark{1}}
\affil{1. Department of Physics and Astronomy, Uppsala University, SE-751 20, Uppsala, Sweden}
\affil{2. Centre for Interdisciplinary Mathematics (CIM), Uppsala University, SE-751 06, Uppsala, Sweden}
\affil{3. Department of Astronomy and The Oskar Klein Centre, Stockholm University, SE-106 91 Stockholm, Sweden}
\affil{4. University of Oulu, Astronomy Research Unit, 90014 Oulu, Finland}



\altaffiltext{2}{beatriz.villarroel@physics.uu.se}
\altaffiltext{3}{anders.nyholm@astro.su.se}

\begin{abstract}

Active Galactic Nuclei (AGN) are extremely powerful cosmic objects, driven by accretion of hot gas upon super-massive black holes. The zoo of AGN classes are divided into two major groups, with Type-1 AGN displaying broad Balmer emission lines and Type-2 narrow ones. For a long time it was believed that a Type-2 AGN is a Type-1 AGN viewed through a dusty kiloparsec-size torus, but an emerging body of observations suggests more than just the viewing angle matters. Here we report significant differences in supernova counts and classes in the first study to date of supernovae near Type-1 and Type-2 AGN host galaxies, using data from the intermediate Palomar Transient Factory, the Sloan Digital Sky Survey Data Release 7 and Galaxy Zoo. We detect many more supernovae in Type-2 AGN hosts (size of effect $\sim$ 5.1$\sigma$) compared to Type-1 hosts, which shows that the two classes of AGN are located inside host galaxies with different properties. In addition, Type-1 and Type-2 AGN that are dominated by star formation according to WISE colours $m_{W1} - m_{W2} < 0.5$ and are matched in 22 $\mu$m absolute magnitude differ by a factor of ten in $L$[OIII]$\lambda$5007 luminosity, suggesting that when residing in similar type of host galaxies Type-1 AGN are much more luminous. Our results demonstrate two more factors that play an important role in completing the current picture: the age of stellar populations and the AGN luminosity. This has immediate consequences for understanding the many AGN classes and galaxy evolution.

\end{abstract}

\keywords{AGN --- supernova --- surveys -- unification -- active -- galactic -- nuclei}

\section{Introduction}\label{sec:intro}

The AGN Unification theory \citep{Antonucci1993} has been subject to many successes and, also, some controversies. In its simplest form only the viewing angle towards the torus matters. This basic picture has long been challenged by statistical tests \citep{Laurikainen,Dultzin,Koulouridis2013}. Many statistical tests are often overlooked due to anisotropic sample selection or small sample sizes \citep{Antonucci2012}. 

Recently, it was shown that even with isotropic selection criteria and large sample sizes of thousands of pairs, the galaxy neighbours to Type-1 and Type-2 AGN differ significantly \citep{VillarroelKorn2014} within a few hundred kiloparsec, including a difference in the number of close neighbour galaxies \citep{Xiaobo}. 

This claim is not uncontroversial. Seemingly, Gordon et al. (2016) do not find any differences in the galaxy neighbours. However, their study suffers from poor statistics (using only tens of pairs) and does not mention some significant colour differences they find in neighbours, see $p$-values from the Kolmogoroff-Smirnoff tests in Table 4 of \cite{Gordon2016}.

Recent works \citep{Donoso2014,Koulouridis2014,Trippe2014,Bitsakis2015} support that factors beyond the viewing angle must play a role \citep{Krongold2002}. Some believe the torus is a nuclear stellar nursery driven by inflows and outflows \citep{Hickups} or the outer parts of a disk-wind coming from the accretion disk \citep{Elitzur2006}, driving an evolutionary sequence from Type-1 $\rightarrow$ Type 1.2/1.5 $\rightarrow$ 1.8/1.9 $\rightarrow$ Type-2 \citep{Elitzur2014}. Presence of spectral features at 10 and 18 $\mu m$ indicate that the AGN has different clumpiness of the torus depending on if it is a Type-1 or Type-2 AGN and if isolated or in merger \citep{Mendoza2015}. The absence of detected broad-line regions in low-luminosity AGN \citep{Nicastro}, and lack of tori at the high-luminosity end, further complicate the picture.

These findings open up for some important questions: what are the true physical differences between the Type-1 and Type-2 AGN populations? Differences could lie in the physics of the central engines, in the structure of the tori \citep{Cristina,Ricci,Elitzur2012}, but also in the host galaxies themselves, e.g. \citet{Heckman1989,Maiolino1995}.

The star-formation histories of the host galaxies can be readily compared with the help of supernovae (SNe), requiring no assumptions about the composition of the galaxy spectra. The luminosity-weighted age of the stellar population is reflected in the occurrence of different SN types: progenitors of core-collapse (c-c) SNe are massive stars ($M \geq 8 ~M_{\odot}$) with short life-times ($< 10^7$ yrs) and indicate recent or ongoing star formation. Thermonuclear SNe, whose progenitors are white dwarfs that take on average $\sim 10^9$ years to form, are indicators of earlier epochs of star formation. Early-type galaxies which are dominated by old stellar populations are not known to host core-collapse SNe, whereas thermonuclear SNe are found in all types of galaxies \citep{Bergh1991}. Recent works indicate a larger fraction of core-collapse to thermonuclear SNe in non-active, star-forming spirals than in spirals hosting AGN. This fraction is connected to the earlier morphological type of the AGN hosts \citep{Hakobyan2014} and the stage of eventual merger in close pairs of galaxies \citep{Hakobyan2013}.

In AGN Unification theory the SN counts and types are expected to be the same for both classes of AGN. In this study using SNe from the intermediate Palomar Transient Factory (iPTF), we test whether Type-1 and Type-2 AGN host galaxies have the same or different occurence rates of SNe. Different SN rates would indicate these AGN types reside in galaxies with different star-formation histories.

The AGN and galaxy samples are taken from the Sloan Digital Sky Survey (SDSS) \citep{York2000} Data Release 7 \citep{Abazajian2009}. The SNe are taken from the iPTF catalogue. This survey is well suited for this work due to its coherent, untargeted mode of detecting transient sources with one and the same telescope.

In Section \ref{sec:methods} we discuss the sample selection and methods. In Section \ref{sec:results} the main results, in Section \ref{sec:Biases} different potential biases that can influence the results. In Section \ref{sec:Statistical} we present the statistical analysis. Finally, we present the conclusions in Section \ref{sec:conclusions}.

\section{Methods}\label{sec:methods}

\subsection{The iPTF survey}\label{sec:iPTF}

The iPTF is an un-targeted wide-field sky survey using the 1.2 m Samuel Oschin telescope (P48) at Palomar Observatory to detect and follow up transient astronomical sources. The scientific scope of the iPTF spans from small solar system objects to extragalactic phenomena. The iPTF project has been running since 2013, and it had a precursor, the Palomar Transient Factory (PTF), which was active during 2009-2012. The technical background of PTF is presented in \citet{Law2009} and the scientific motivation in \citet{Rau2009}. The iPTF is presented in \citet{Kulkarni2013}. For brevity, we will refer to our SN catalogue (including SNe found during the PTF period) simply as the iPTF catalogue.

In this study, we use the iPTF SN catalogue from the time window between 2009 March 2 until 2014 June 17. This catalogue contains 2190 extragalactic SNe with known right ascension ($\alpha$), declination ($\delta$), spectroscopic redshift ($z_{SN}$) and spectroscopic classification. This SN catalogue contains 1494 thermonuclear SNe (i.e. Type Ia) and 632 core-collapse SNe. In the core-collapse category, we include SNe Types Ib, Ic, Ib/c, Ibn, II, IIP, IIL, IIn and IIb. The remaining 64 SNe are either of unclear spectroscopic type or superluminous SNe. None of the 11 superluminous SNe with $z < 0.2$ in our catalogue was found in a galaxy with a spectrum in SDSS DR7. The superluminous SNe can therefore, for our purposes, be included in the remainder category.

The SNe in this sample are located at declinations $-25^{\circ} < \delta < 80^{\circ}$, mostly at galactic latitudes $|b| > 20^{\circ}$. The mean redshift of the sample is $z \approx 0.09$, with 95 \% of the SNe having $z < 0.2$. For a motivation of curtailing the SN catalogue at 2014 June 17, see Sect. \ref{sec:SNbiasiPTF}.

An advantage of the iPTF catalogue is its cohesive nature. All the SNe have been discovered during an untargeted search with the same telescope, and spectroscopic classifications have been made in a timely fashion. The compatibility with SDSS in sky coverage makes the iPTF SNe sample suitable for our investigation. Our SDSS samples covers $-10^{\circ} <\delta < 70^{\circ}$ and galactic latitudes $|b| > 20^{\circ}$, comparable to the distribution of the SN locations.

\subsection{Comparison with other supernova catalogues}\label{sec:OtherCatalogues}

Another SN catalogue that comes to mind is the SDSS SN catalogue \citep{Sako2014}. The 902 confirmed SNe in the SDSS SN survey\footnote{Listed at \url{http://classic.sdss.org/supernova/snlist.dat}} represents a SN sample of smaller scope than the iPTF sample. Not all SNe are spectroscopically classified, but the survey is deep and homogenous. Unfortunately, the SNe are all located in the SDSS southern equatorial Stripe 82, which is outside the area of the SDSS DR7 sample from the central region used in this study. This renders the SDSS SNe unsuitable for our purposes.

The Asiago SN catalogue \citep{Barbon1999} is a historically comprehensive compilation of extragalactic SNe. The catalogue encompasses 6530 SNe as of 2016 March 27, in both hemispheres. The Asiago catalogue is extensive but has uneven quality -- some of the SNe in it lack spectroscopic classification, some lack spectroscopic redshift. The SN catalogue compiled by \citet{Lennarz2012} contains data for 5526 extragalactic SNe, whereas the Sternberg SN catalogue, presented by \citet{Tsvetkov2004}, contains less than 3000 extragalactic SNe. The circumstance that they are compiled from a wide range of sources make them less suitable compared to the cohesive iPTF catalogue. The same holds for The Open Supernova Catalogue \citep{guillochon16}, with $\approx 37000$ SNe (of which 12 \% have spectra in the catalogue) as of 2016 December, collected from different public sources.

As discussed by \citet{Anderson2013}, earlier SN searches have prioritised SN detection over completeness with respect to SN types or host galaxy types. This bias should affect such compilations as the Asiago catalogue and other catalogues listing SNe found before the start of untargeted SN searches.

\subsection{Selection of AGN}\label{sec:SelectionAGN}

The samples of host galaxies were obtained through the SDSS Data Release 7. We select objects classified as 
either 'Quasars' or 'Galaxies', within redshift $0.03 < z < 0.2$, unless flagged for brightness (flags\&0x2=0), saturation (flags\&0x40000=0), or blending (flags\&0x8=0) \citep[their table 9]{Stoughton02}.

The emission lines are obtained from the SpecLine table in DR7. We require that the objects have H$\alpha$ in emission and select Type-1 AGN, Type-2 AGN and star-forming galaxies using optical emission line diagnostics. Our Type-1 AGN are objects with $\sigma$(H$\alpha$) $>$10 \AA\ (or FWHM(H$\alpha$) $>$ 1000 km/s). The Type-2 AGN have narrow lines $\sigma$(H$\alpha$) $<$ 10 \AA\, fulfilling the Kauffmann criterion \citep{Kauffmann2003}:

\begin{equation}
\log([\ion{O}{iii}]/{\rm H}\beta) > 0.61/(\log([\ion{N}{ii}]/{\rm H}\alpha))-0.05)+1.3 \label{eq,BPT}
\end{equation}\label{BPTequation}

The star-forming galaxies are defined as all the other narrow-line objects.

In this way, we classify the objects into Type-1s, Type-2s  and star-forming galaxies, referred to as ``largest samples'', using optical emission line diagnostics. For all objects we search for morphological classifications ('Spiral', 'Elliptical', 'Uncertain') from the project Galaxy Zoo 1 \citep{Lintott,Lintott2010}, emission line measurements, redshifts and celestial coordinates, leaving ``parent samples'' of 11632 Type-1 AGN (1864 spiral), 77708 Type-2 AGN (36720 spiral) and 137489 star-forming galaxies (49072 spiral). For the vast majority of these objects we can also find Wide-field Infrared Survey Explorer (WISE) magnitudes.

\subsubsection{Refined samples}\label{sec:refinedsamples}

As additional samples later used only for direct comparison of host galaxy properties, we also create some 'refined samples'. Starting from the parent samples, we select only face-on spiral hosts with high S/N in the emission lines from Galaxy Zoo Data Release 2 \citep{Willett}, minimizing dust extinction due to host galaxy inclination. We also require $S/N > $ 3 in H$\alpha$, minimum SDSS Gaussian line heights $h$(H$\alpha$) $>$ 10 * 10$^{-17} $erg/s/cm$^{2}$/\AA\ and h(H$\beta$) $>$ 5 * 10$^{-17} $erg/s/cm$^{2}$/\AA\, in order to avoid effects of stellar absorption affecting weak lines in our classification. Out of these, we select only those having WISE colours.

The refined samples will be used for comparing star formation with WISE colours in Type-1 and Type-2 AGN or the [\ion{O}{iii}]5007 in host galaxies matched by amount of dust.

\subsection{Pairwise matching}\label{sec:Pairwise}

The numbers of coherently collected SNe are scarce (2190 in our sample as on 2014 June 17). Thus, we create pairwise matched subsamples of Type-1 AGN, Type-2 AGN and  star-forming objects to compare objects as similar to each other as possible in redshift distribution and selected properties e.g. the luminosity $L$[\ion{O}{iii}]5007. We compare (i) Type-1 AGN to Type-2 AGN, and (ii) Type-2 AGN to star-forming galaxies. The aim with the latter test is to probe whether the observed Type-2 AGN properties can be explained by star formation alone.

For two samples of intrinsically similar objects the probability of detecting faint SNe is the same if they have similar redshift distributions. Therefore, for each galaxy in the parent sample, we select a galaxy from the second parent sample having the closest value in redshift and a specific property of interest. After the matching is done, we first throw away all matched pairs that differ more than 20\% in the property of interest.

Four types of specific properties and matchings are explored:

\begin{enumerate}

\item Redshift only. This allows to remove biases in Galaxy Zoo morphology classifications as well as the Malmquist bias.

\item $L$[\ion{O}{iii}]5007 from the narrow-line region (NLR). In the simplest AGN Unification, the NLR is believed to be isotropically distributed outside the torus and to be equally strong for AGN of the same activity level irrespective of the viewing angle. Selecting on $L$[\ion{O}{iii}]5007 -- meaning one selects all Type-1 and Type-2 AGN above a selected certain line flux in a sample -- should give the same host galaxy properties under the conditions of isotropy \citep{Antonucci1993}. Matching on $L$[\ion{O}{iii}]5007 is similar to selecting on $L$[\ion{O}{iii}]5007 if the two $L$[\ion{O}{iii}]5007 distributions are the same (as predicted by the simplest Unification) but can be problematic if the distributions differ at the high-luminosity end. Moreover, we expect the same line width $\sigma$[\ion{O}{iii}]5007 in matched samples Type-1 and Type-2 AGN samples.

\item WISE $M_{w4}$ (22$\mu m$) absolute magnitude. We use this match as our stellar-mass proxy assuming the dust emission traces the stellar mass. The 22$\mu m$ magnitude is a good measure of heated-dust emission in the host galaxy, especially in galaxies where the torus contribution to the total 22$\mu m$ is negligible in comparison, meaning galaxies having $m_{W1} - m_{W2} < 0.5$ \citep{Wright}. A less favourable option is to use dust reddening $F$(H$\alpha$/H$\beta$), but if dust reddening in the BLR and NLR differs \citep{Gaskell}, the matching will be biased. We therefore neither correct $L$[\ion{O}{iii}]5007 for dust reddening. This matching is only done for the galaxies in the parent samples that have WISE magnitudes.

\item Exponential fit scale radius (r-band). As the star-formation history depends on the available gas mass, matching by an apparent measure of the galaxy volume should minimize differences in star-formation histories under the assumption of a mass-size relation.

\end{enumerate}

We do a two-sample Kolmogoroff-Smirnoff test for each matched property to ensure the sample distributions are statistically similar. For the $M_{w4}$-matched samples we had to do a finer match by throwing away matched pairs differing more than 5\% in $M_{w4}$.

\begin{figure*}
 \centering
   \includegraphics[scale=.8]{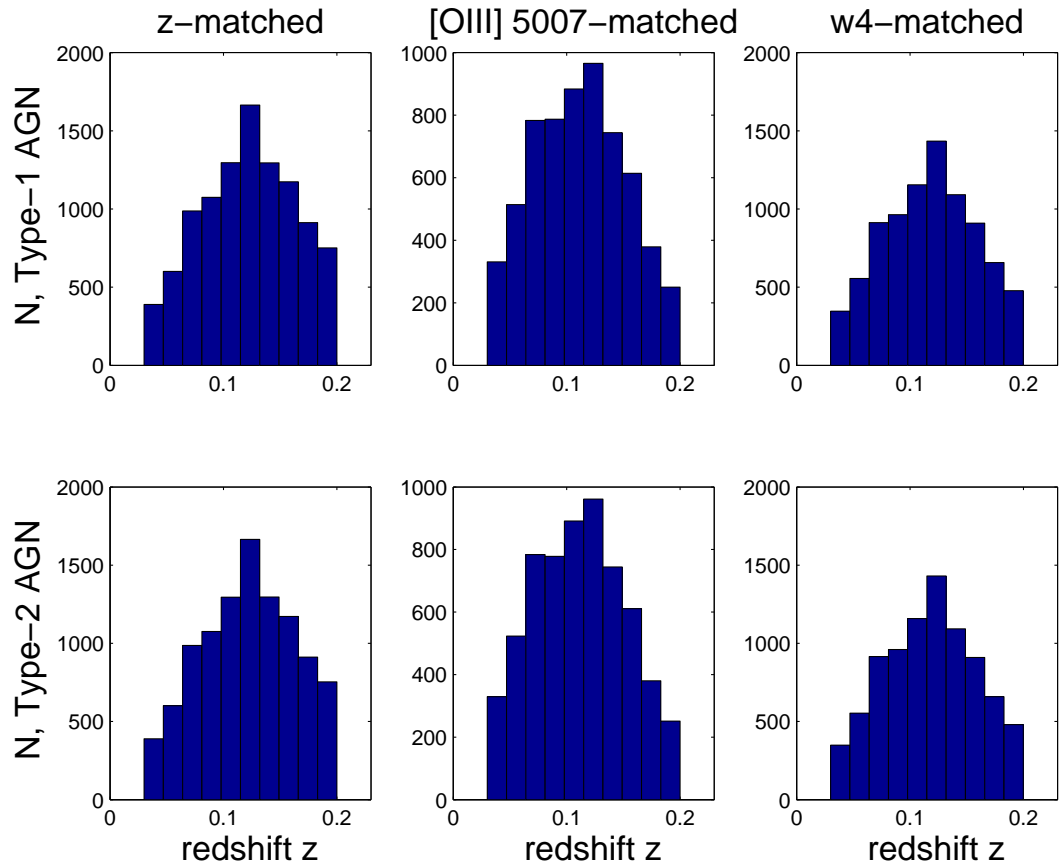}
     \caption{The redshift distributions of matched samples. The redshift distributions of matched Type-1 and Type-2 AGN are demonstrated for three different types of matchings: redshift, $L$[\ion{O}{iii}]5007 and $m_{w4}$. Two-sample Kolmogoroff-Smirnoff tests confirm the pairwise matched Type-1 and Type-2 distributions are the same.}
               \label{RedshiftDistributions}
     \end{figure*}

\begin{figure*}
 \centering
   \includegraphics[scale=.8]{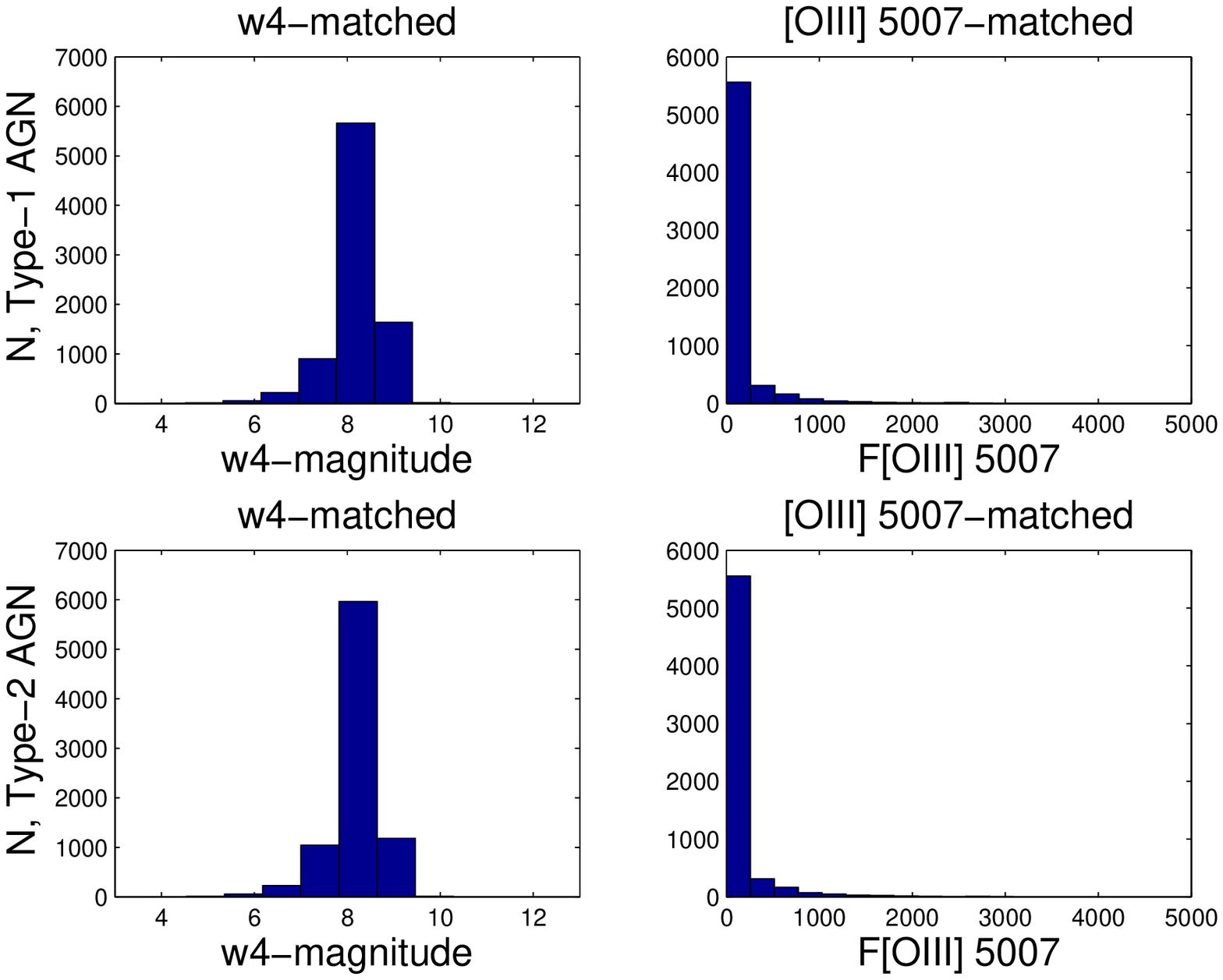}
     \caption{The distributions of $L$[\ion{O}{iii}]5007 and $m_{w4}$. The distributions of $L$[\ion{O}{iii}]5007 or $m_{w4}$ in matched Type-1 and Type-2 AGN samples are demonstrated for two different types of matchings: $L$[\ion{O}{iii}]5007 and 
$m_{w4}$. In the $L$[\ion{O}{iii}]5007 histograms, four objects in each sample above $F > $ 5000 are not plotted in the histogram due to visibility. Two-sample Kolmogoroff-Smirnoff tests confirm the pairwise matched Type-1 and Type-2 distributions are the same.}
               \label{OtherDistributions}
     \end{figure*}

The pairwise matched subsamples for examining SN counts are created by matching in three different parameters (redshift $z$, luminosity $L$[\ion{O}{iii}]5007 and $m_{W4}$) as described earlier. The sizes of the pairwise matched subsamples can be found in Table \ref{SupernovaAll}. An example of the redshift and property distributions in the matched Type-1 and Type-2 samples can be seen in Figure \ref{RedshiftDistributions} and \ref{OtherDistributions}. The similarity of all the matched samples is ensured through two-sample Kolmogoroff-Smirnoff where the null hypothesis (that two matched samples are similar) holds for the nominal value $\alpha$=0.05. The procedure is repeated for pairwise matched Type-2 AGN and star-forming galaxies.

Finally, the refined samples of face-on hosts with high signal-to-noise are also matched with the same method. In addition, the refined samples
are matched by Balmer decrement $F$(H$\alpha$)/$F$(H$\beta$).

\subsection{Matching of supernovae}\label{sec:SNmatching}

A SN is considered to be matched with a galaxy if the following two conditions are both satisfied:

\begin{itemize}

\item The projected distance on the plane of the sky between the SN and the galaxy is computed (plane approximation). The search radius $d$ is set by converting the desired physical search radius into a angular radius at a galaxy distance assumed to be $z_{AGN} \cdot c / h_0$. A Hubble constant ${h_0}$ = 72 $\rm km$ $\rm s^{-1}$ $\rm Mpc^{-1}$ is used throughout this work. If the SN is found to lie inside the given search radius, the compliance with a redshift condition is also checked.

\item Our redshift condition is $|z_{SN} - z_{AGN}| < 0.003$ and accounts for redshifts due to peculiar motions. This is more strict than the redshift matching condition of $|z_{SN} - z_{AGN}| < 0.01$ used in the study by \citet{Wang2010}. In their study, however, most (97 \%) of their sample of 620 cross-matches fulfills the $|z_{SN} - z_{AGN}| < 0.003$ condition.

\end{itemize}

\begin{figure*}
 \centering
   \includegraphics[scale=.8]{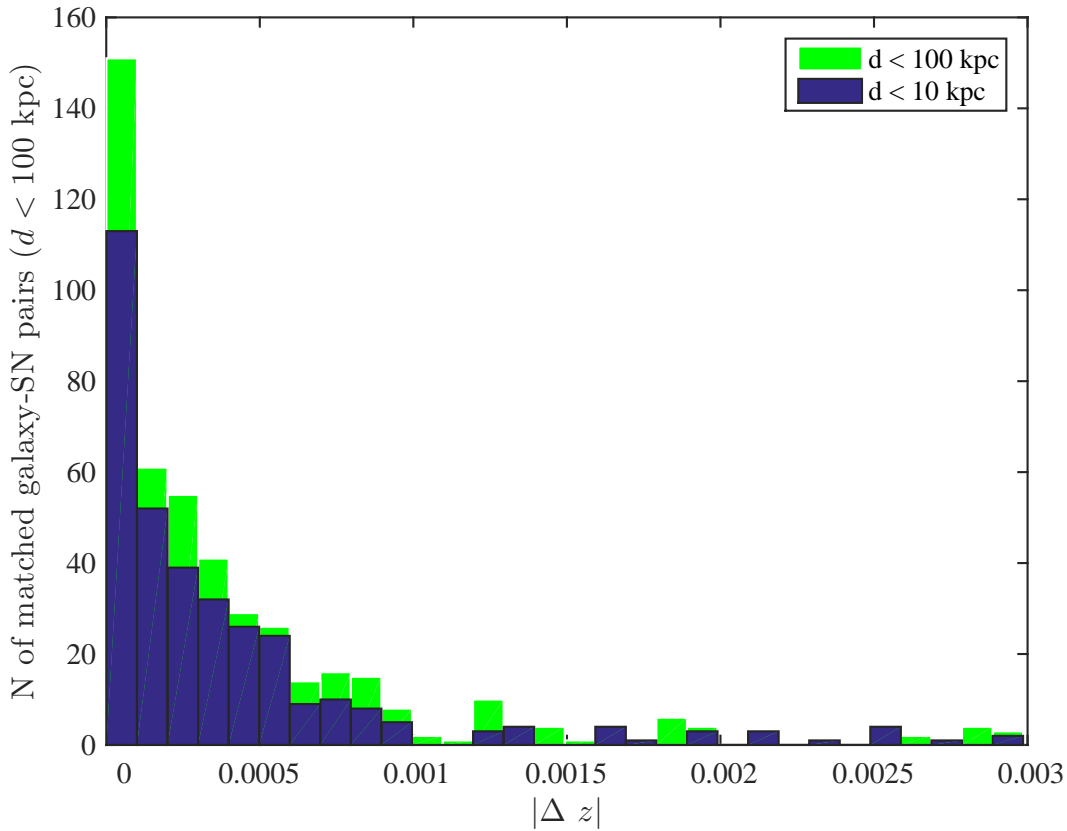}
     \caption{The $|\Delta z|$=$|z_{SN} - z_{AGN}|$ distribution for SNe near host galaxies in the largest samples. The figure includes all SNe found around the galaxies fulfilling the $|z_{SN} - z_{AGN}| < 0.003$ and the distance criterions $d < 10$ kpc (blue) and $d < 100$ kpc (green). The $|\Delta z|$ criterion is selected as the host galaxy redshift might be influenced by its peculiar motion. In a few cases, the redshift error of the SN is larger (up to $\delta z <$ 0.005), if the redshift determination is done directly from the SN itself as effects from SN expansion velocities and Doppler broadening enter \citep{Blondin2007}. Out of the total 471 galaxy-SN pairs, 416 (88 \%) are located within $|z_{SN} - z_{AGN}| < 0.001$. The bottom-heavy distribution indicates a physical association, on a population level, between the host galaxies and the SNe. We matched 215 individual SNe, showing that the total number of 471 matched galaxy-SN pairs depends on the occurrence of the same galaxy in several different samples.}
     \label{SupernovaGalaxyDelta}
     \end{figure*}

If both these conditions are fulfilled, the SN is considered to be associated with the galaxy. The SNe inside a galaxy are collected by the $d < $ 10 kpc, and those inside a galaxy or in a close companion are found with the $d < $ 100 kpc criterion. If this is wrong and there is no association between the AGN and the SN, then the distribution of different SNe with different $|\Delta z|$ = $|z_{SN} - z_{AGN}|$ ought to be a uniform distribution. This can be checked by plotting a histogram with the $|\Delta z|$ between the AGN and the SNe, see Fig \ref{SupernovaGalaxyDelta}. It is clear that the distribution is bottom-heavy and that most of the SNe are associated with the host galaxies.

It is important to point out, that the SN redshift determination method brings in uncertainty regarding the redshift uncertainties. Most of the time the redshift is measured from the host galaxy of the SN and is rather accurate. But sometimes, the SN redshift is measured directly from the SN itself and influenced by Doppler broadening and expansion velocities, being slightly larger \citep{Blondin2007}.

The degree of association between SNe and galaxies can also be checked by visually inspecting the SDSS DR7 images of the galaxies, with the SN positions overplotted. We note the subjectivity involved, e.g. in cases where a SN occurs in an interacting pair of galaxies. For the largest sample of Type-2 AGNs (77708 galaxies), and the 59 SNe matched to them, about 10 \% of the SNe visually appears to be located in a companion galaxy of a sample galaxy. For the largest sample of starforming galaxies (137489 galaxies), and the 152 SNe matched to them, a comparable fraction (about 9 \%) appears to be located in a companion galaxy. These results from visual inspection of SN positions complements the conclusions drawn from Fig. \ref{SupernovaGalaxyDelta}.

\section{Results}\label{sec:results}

\subsection{Supernova counts}

We begin our analysis by counting SNe within two different projected distances from the center  of the host galaxies: 10 kpc (within the typical radius of a spiral galaxy),  100 kpc (within the galaxy or a possible close companion). We start with the largest samples, see Table \ref{Largest}.

The first thing to notice is the lack of SNe around Type-1 AGN at projected separations d $<$ 100 kpc. Only one SN is found. The number of SNe around Type-2 AGN are clearly systematically higher for the largest samples in Table \ref{Largest} (size of effect $\sim$ 5.1$\sigma$). We verify that this is no bias due to potential difficulties in detecting SN in the immediate vicinity of bright or transient AGN by excluding all SNe within 3 arcseconds from the host centres, see Section 4.3.2. This still yields 1 SN near 11632 Type-1 AGN (detection fraction $\sim$ 8.6 $\times$ 10$^{-5}$) and 46 SN near 77708 Type-2 AGN (detection fraction $\sim$ 5.9 $\times$ 10$^{-4}$) -- a significant difference (size of effect $\sim $4.1 $\sigma$, see Section \ref{sec:unc}). Visual inspection show that the majority of the SNe ($>$ 90 \%) come directly from the AGN hosts (only a small fraction from the companion galaxies).

For the Type-1 and Type-2 samples matched in redshift or $M_{w4}$ the differences are significant. But for the $L$[\ion{O}{iii}]5007-matched samples we only find one SN in each sample, yielding no difference at all. Does this mean the Simplest Unification is valid and there is no difference in galaxy properties and SN counts in samples selected and matched in $L$[\ion{O}{iii}]5007? If so, also the host morphologies for the matched samples should be the same. But the fraction of Type-1 and Type-2 AGN in spiral hosts is 20 \% vs 44 \%, in disagreement with the Simplest Unification. This suggests that the lack of significant difference in SN counts stems from poor statistics.

As an alternative test, we select on $L$[\ion{O}{iii}]5007. This should be suitable as in the Simplest Unification no difference is expected in $L$[\ion{O}{iii}]5007 on the higher luminosity-end (while matching removes potential differences at the high-luminosity end). We do this twice for the largest samples using both $F$[\ion{O}{iii}]5007 $> 10$ $\times$ 10$^{-17}$ erg/s/cm$^{2}$ and $F$[\ion{O}{iii}]5007 $> 30$ $\times$ 10$^{-17}$ erg/s/cm$^{2}$ as flux limits. The difference in $L$[\ion{O}{iii}]5007-selected Type-1 and Type-2 AGN samples is noteable ($F$[\ion{O}{iii}]5007 $ > 10$ case: size of effect $\sim $2.8 $\sigma$, $F$[\ion{O}{iii}]5007 $> 30$ case: 1.8 $\sigma$) following the same trend of Table \ref{Largest}. This reflects purely changes in the sample sizes, unless the difference in matched vs selected samples originates in a break-down of Unification at the higher luminosities.

As early-type objects are well-known \citep{Li2011} for having few SNe setting off, it would be ideal to use only spiral hosts. We do not find a single SN around spiral Type-1 AGN. But using a statistical hypothesis test \citep{Krishnamoorthy} (see Section \ref{sec:pairwise}) only a borderline-significant difference in SN counts for Type-1 and Type-2 AGN at $d < $ 100 kpc (p-value $\sim$ 0.06) is found, showing the need of larger SN samples.

\begin{figure*}
 \centering
   \includegraphics[scale=.5]{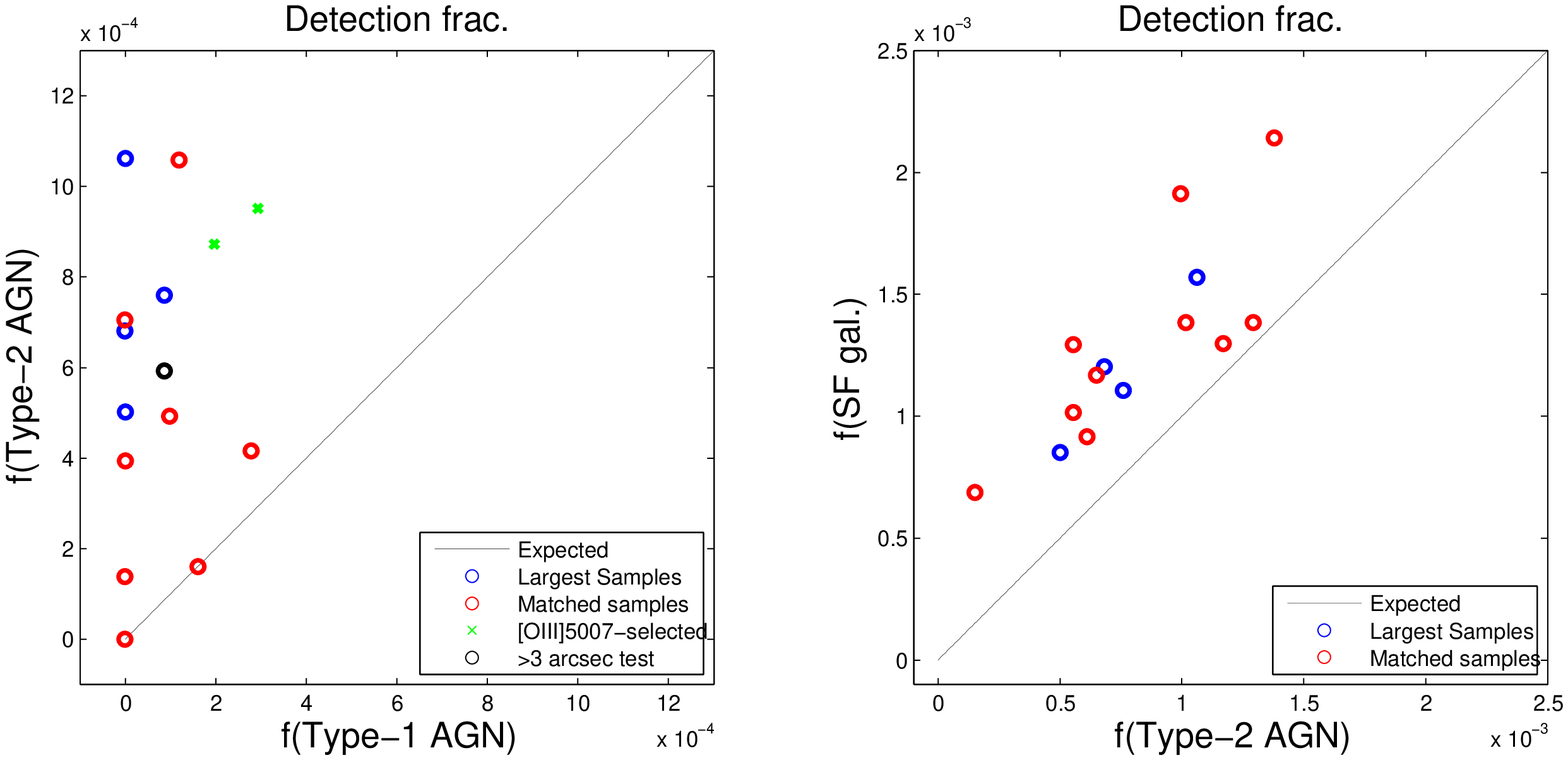}
     \caption{Visual presentation of collected SN counts. For each sample used in the paper, the detection fraction of galaxies showing a SN (``f(type)'') in the samples are plotted against each other. In this figure, the points are jittered to reduce overlaps of scatter points. As the samples are overlapping, the points are not independent of each other. In the left panel, results from Table \ref{Largest}, Table \ref{SupernovaAll}, the $L$[\ion{O}{iii}]5007-selected samples and the r $>$ 3 arcsec-test. The dark-grey line at the left represents expectations from the simplest Unification. In the left plot, only the $L$[\ion{O}{iii}]5007-matched samples fall upon the line (the two red circles at the left bottom), while the $L$[\ion{O}{iii}]5007-selected samples (black circles) end up above. Furthermore, if the only difference between Type-2 and Type-1 AGN was the amount of dust (Type-2 being dustier), the points would end up $below$ the grey line. This is opposite from what is seen. In the right panel, the SN counts are shown for Type-2 vs star-forming galaxies in spiral hosts. The points follow the shape of the grey line, as expected if the two types of host galaxies are the same.}
               \label{SupernovaCountsPlot}
     \end{figure*}

For the pairwise matched samples the counts are too small to show any significance individually. The $L$[\ion{O}{iii}]5007-matched samples we commented upon earlier, while for the redshift-matched samples a significant difference is found for $d < $ 10 kpc. The lack of significant difference for $d < $ 100 kpc therefore stems from the poor statistics. The samples matched in redshift and apparent galaxy size yield small, insignificant differences between Type-1 and Type-2 AGN. As the size of the galaxy depends on the star-formation history, no difference in counts could mean that we have either matched successfully in the star-formation history, or more realistically, that the sample sizes are too small. However, the collected results in Table \ref{SupernovaAll} are visually presented in Figure \ref{SupernovaCountsPlot} where the left column reinforces our earlier conclusions: Type-2 AGN hosts have higher SN rates than Type-1 AGN hosts, much higher than from the expectations of the simplest Unification theory (represented by the grey line). Also the insignificant differences fall into the same area of the plot.

The larger count of SNe could mean either a difference in stellar age or stellar mass (or both). But samples unmatched in stellar mass lead to differences in clustering on Mpc scale \citep{Mendez} while we find no differences in SN counts on large scale, see Sections \ref{sec:largescale1} and \ref{sec:largescale2}. Moreover, the matching in $M_{w4}$ -- our proxy for stellar mass -- still yields significant different SN counts. This suggests that the discrepancy in SN counts is due to differences in stellar age between Type-1 and Type-2 AGN.

The conclusion is supported by the much higher $w2-w3$ colour indices in Type-2s indicating stronger star formation \citep{Coziol2015} in samples consisting of face-on, spiral-host Type-1 and Type-2 AGN that are dominated by dust emission from stars with $m_{W1} - m_{W2} < 0.5$ \citep{Wright} and pair-wise matched in $L$[\ion{O}{iii}]5007, see Section \ref{sec:AltSFR}. 

Less striking differences are found between Type-2 AGN and star-forming galaxies. Star-forming spirals show higher core-collapse SN counts at d $<$ 10 kpc ($\sim$ 3.3$\sigma$) and even in the redshift-matched samples a difference is present. The right column in Figure \ref{SupernovaCountsPlot} shows a relationship between the two classes, although offset in y-axis following the shape of the grey line.

\subsection{Anisotropy?}

One may argue that the $L$[\ion{O}{iii}]5007 may be anisotropic or even have different physical origins in Type-1 and Type-2 AGN. If so, one expects to find differences in the Gaussian line width of the 
[\ion{O}{iii}]5007 emission in Type-1s and Type-2s and/or differences in SN counts.

\subsubsection{SN samples: anisotropic sample selection?}\label{sec:AnisotropicSN}

As Type-1 AGN have a contribution of a non-stellar, power-law continuum component to their observed luminosity, this could influence the detections of Type-1 and Type-2 AGN. However, the emission lines are independent of the continuum emission and in the Simplest Unification $L$[\ion{O}{iii}]5007 is expected to be an isotropic indicator of AGN luminosity.

Type-1 and Type-2 samples selected on, or matched in, $L$[\ion{O}{iii}]5007 should be the same in all other properties: they should have similar host galaxy types and similar NLR kinematics. However, in \citep{VillarroelKorn2014} $L$[\ion{O}{iii}]5007-matched hosts showed different colours of their galaxy neighbours.

Using $L$[\ion{O}{iii}]5007-matching we see no difference, or cannot see, in the SN counts near Type-1 and Type-2 AGN: we find one SN near Type-1 hosts and one SN near Type-2 hosts.

While matching is good, a better way is still to select on $F$[\ion{O}{iii}]5007. Selecting on $F$[\ion{O}{iii}]5007 means that all Type-1 and Type-2 above a certain flux value in $F$[\ion{O}{iii}]5007 should have the same properties. We try this for two separate lower flux limits: $F > $ 10 or 30 $\times$ 10$^{-17}$ erg/s/cm$^{2}$, yielding the following SN counts within $d <$ 100 kpc from the host galaxies in the largest samples:

\begin{enumerate}

\item $F > $ 10 $\times $10$^{-17}$ erg/s/cm$^{2}$. We have 5101 Type-1 AGN and 47010 Type-2 AGN. The number of SNe near Type-1 AGN is 1 (detection fraction $\sim$ 1.96 $\ast$ 10$^{-4}$) The number of SNe near Type-2 AGN is 41 (detection fraction $\sim$ 8.72 $\ast$ 10$^{-4}$). The estimate of the size of effect is $\sim$ 2.8$\sigma$.

\item $F > $ 30 $\times $10$^{-17}$ erg/s/cm$^{2}$: We have 3403 Type-1 AGN and 19976 Type-2 AGN. The number of SNe near Type-1 AGN is 1 (detection fraction $\sim$ 2.94 $\ast$ 10$^{-4}$). The number of SNe near Type-2 AGN is 19 (detection fraction $\sim$ 9.51 $\ast$ 10$^{-4}$). The estimate of the size of effect is $\sim$ 1.8$\sigma$. 

\end{enumerate}

The number of SNe in the largest Type-2 host samples (77709 objects) within $d < $ 100 kpc was 59 SNe (see Table \ref{Largest}). Considering the smaller size of the flux-restricted Type-2 AGN samples, the expected new SN counts are: (a) $F > $ 10: 47010/77709*59 SNe $\sim$ 35, (b) 19976/77709*59 $\sim$ 15. Therefore, we can easily see that the loss of significance in the $F$[\ion{O}{iii}]5007-selected sample reflects upon the decreased number of objects in the samples.

\subsubsection{NLR kinematics: anisotropic sample selection?}\label{sec:AnisotropicNLR}

To further probe the relevant NLR kinematics behind the $L$[\ion{O}{iii}]5007 emission, we analyze the AGN themselves. If the same physical mechanisms are causing the [\ion{O}{iii}]5007 emission in Type-1s and Type-2s -- and isotropically -- the [\ion{O}{iii}]5007 luminosity-normalized line widths $\sigma$([\ion{O}{iii}]5007) must be the same. 

When we explore the Gaussian line widths $\sigma$([\ion{O}{iii}]5007) of the various classes of objects, we use refined samples matched in redshift and $F$[\ion{O}{iii}]5007.

The normally distributed $\log$ of the $\sigma$([\ion{O}{iii}]5007) values for estimating means and errors are calculated for $L$[\ion{O}{iii}]5007-matched, face-on spiral hosts. No significant difference in line width for the Type-1s and Type-2 objects is found ($p=$ 0.13). This disagrees with the earlier observation that the NLR has a component of motion giving rise to geometric differences \citep{GaskellOutflow}. A clearer line width difference between the Type-2 ($\log_{10}$($\sigma$)= 0.242 $\pm$ 0.003) and star-forming objects ($\log_{10}$($\sigma$)=0.191 $\pm$ 0.004). The slighty wider [\ion{O}{iii}]5007 line in the Type-1 and Type-2 AGN (over star-forming galaxies) support that a significant contribution to the [\ion{O}{iii}]5007 flux might arise isotropically distributed in a region close to the AGN nucleus where the clouds rotate around the center of the galaxy at higher velocities, causing additional Doppler broadening. But also outflows in AGN are known to cause broadening of the [\ion{O}{iii}]5007 line and can at high redshift $z \sim$ 2.5 in extreme cases show broadening corresponding to $2600 - 5000$ km/s \citep{Zakamska2016}. Perhaps, the slight differences in [\ion{O}{iii}]5007 between the AGN and star-forming galaxies therefore may indicate the presence of outflows from the nucleus.

The results are displayed in Table \ref{MatchedZooCorrect}.

\begin{table*}[ht]
\caption{$\sigma$([\ion{O}{iii}]5007) in the refined [\ion{O}{iii}]5007 pairwise matched samples.
The errors indicate standard errors assuming Gaussianity of $log10$($\sigma$). The reference samples are marked by star '*'.}
\centering
\begin{tabular}{c c c}
\hline\hline
\multicolumn{3}{c}{Samples} \\
\hline\hline
Type of match, ``Unification''& $N$ & $log10$($\sigma$)\\ 
Type-1* & 123 & 0.4713 $\pm$ 0.0140\\
Type-2 & 123 & 0.4377 $\pm$ 0.0101\\[0.2ex]
\hline
Type of match, ``Stars''& $N$ & $log10$($\sigma$)\\ [0.2ex]
\hline
Type-2* & 1244 & 0.2415 $\pm$ 0.0035\\
Star-forming & 1244 & 0.1906 $\pm$ 0.0036\\[0.2ex]
\hline
\end{tabular}
\label{MatchedZooCorrect}
\end{table*}

\subsection{AGN luminosity}

If the obscuration is the dominant factor that separates Type-2 from Type-1 AGN, one may expect that the refined samples of Type-1s and Type-2s matched in the heated-dust emission from their host galaxy are more or less as luminous in $L$[\ion{O}{iii}]5007. For the 137 paired Type-1s and Type-2s in the refined samples matched in redshift and $M_{w4}$, we compare the mean $L$[\ion{O}{iii}]5007. The mean $\log_{10}$$L$[\ion{O}{iii}]5007 [erg/s] is 40.934 $\pm$ 0.0455 for Type-1 objects and 39.9738 $\pm$ 0.0889 for Type-2s. This demonstrates that Type-1s are much more luminous ($\sim$ 9.6$\sigma$) than Type-2s in host galaxies with similar dust distributions. The effect is equally convincing if exploring and matching objects with WISE bands dominated by star formation $m_{W1} - m_{W2} < 0.5$.

\subsubsection{Refined samples dominated by dust emission near stars}\label{sec:DustEmission}

With the simple criterion $w1 -w2 >= 0.8$ one can easily identify sources dominated by AGN \citep{Assef}. In general, $WISE$-selected samples are biased towards more Seyfert-like AGN with low luminosity \citep{Mingo}. We do this for our Type-1 and Type-2 AGN and find that majority in the matched samples have $w1 -w2 < $ 0.8, i.e. have the WISE bands dominated by dust heated by star-formation, permitting the comparison using $w2 - w3$ colours. An example are the $L$[\ion{O}{iii}]5007-matched samples where only 109 out of 123 Type-1 AGN and 116 out of 123 Type-2 AGN have their dust dominantly heated by star-formation.

The best way to find out if this influenced our conclusions, is by redoing the tests using only objects that fulfil the $w1 - w2$ $<$ 0.8 condition by Assef et al. (2012) and, or the even the stricter Wright et al. (2010) condition $w1 - w2$ $<$ 0.5. Using these criteria the objects clearly are dominated by dust emission near the stars and comparison of the $w2 - w3$ colour as well as the $w4$-matching is valid. These objects we match by either $w4$ or $L$[\ion{O}{iii}]5007. The resulting samples are clearly smaller.

The resulting, new $w4$- and $L$[\ion{O}{iii}]5007-matched samples show the same average $w2 - w3$ as before. Type-2 shows again higher star-formation than Type-1. Also here in the $w4$-matched samples one can see that Type-1 are much luminous than Type-2 AGN. An alternative view can be gained from comparing the $M_{w4}$ in $L$[\ion{O}{iii}]5007-matched samples: -30.2 +/- 0.1 mag (Type-1 AGN) or -30.8 +/- 0.1 mag (Type-2 AGN), supporting there is more dust in Type-2 AGN hosts than in Type-1 AGN hosts.

One may wonder if this particular result has any connection to the receding torus model \citep{Lawrence}, where the opening angle of the torus gets larger with increasing AGN luminosity. The increased ratio of Type-1/Type-2 AGN at larger luminosities \citep{Simpson2005,Lusso2013} supports the idea of a receding torus. However, as it appears that the age of the stellar population differs between Type-1 and Type-2 AGN, it seems more reasonable that the difference in the dust is on host galaxy scale. While our results say that Type-1 AGN are more luminous than Type-2 AGN, they do not support a receding torus per se.

\section{Biases}\label{sec:Biases}

There are some potential selection biases that can influence our samples. Weak lines used in the object classification might be influenced by stellar absorption. In this study, demanding a $S/N > $ 3 in H$\alpha$ emission before selection gives similar results on the SN counts in Table \ref{Largest} \& \ref{SupernovaAll}, but with poorer statistics. Dust extinction effects due to inclination might potentially influence the emission line strengths \citep{Baker1997} or the SN detection rate. The differences in how Type-1 and Type-2 AGN are classified (Type-1 needs only emission in H$\alpha$, while Type-2 AGN in three additional lines) can bias the Type-2 AGN towards more star-forming hosts. But also might the selection of Type-1s and Type-2s become rather anisotropic due to the contribution of a non-stellar continuum component to the observed luminosity of Type-1 AGN.

Many of these issues where treated in the close neighbours study \citep{VillarroelKorn2014}, correcting for these typical biases e.g. increasing the $S/N$ ratio of the used emission lines also did not change the results in any way. There, also effects from the removal of LINERs and the clumpy tori were explored, showing no influence on the outcome.

Nevertheless, we here present tests dealing with some particular problems related to the current study.

\subsection{Star formation in the refined samples}\label{sec:AltSFR}

To control several biases at the same time, we use the refined, pairwise matched subsamples. The pairwise matching ensures our samples to have the same redshift distribution and also the same distribution in one of the remaining four parameters separately (only $z$, $L$[\ion{O}{iii}]5007, $F$(H$\alpha$/H$\beta$), $m_{w4}$). The refined sample sizes are shown in Table \ref{WISERefinedZooSamples}.

\begin{table*}[t]
\caption{Comparing average $w2 - w3$ colours of objects in the refined pair-wise matched samples of face-on, spiral hosts. The errors indicated are standard errors, assuming Gaussianity of the underlying colour distribution. The reference sample is marked by star '*'.}
\centering
\begin{tabular}{|c | c c c c|}
\hline\hline
\multicolumn{5}{c}{Type-1* vs Type-2} \\
\hline\hline
Type of match & $N$ Type-1* & $N$ Type-2 & Type-1, $w2 - w3$ & Type-2, $w2 - w3$ \\
$z$ & 178 & 178 & 2.976 $\pm$ 0.02 & 3.487 $\pm$ 0.03\\
$L$[\ion{O}{iii}]5007 & 123 & 123 & 2.938 $\pm$ 0.030 & 3.489 $\pm$ 0.04\\
$F$(H$\alpha$/H$\beta$) & 136 & 136 & 2.981 $\pm$ 0.030 & 3.496 $\pm$ 0.03\\
$m_{w4}$ & 176 & 176 & 2.975 $\pm$ 0.02 & 3.672 $\pm$ 0.03\\[0.2ex]
\hline
\multicolumn{5}{c}{Type-2* vs star-forming} \\
\hline
Type of match & $N$ Type-1* & $N$ Type-2 & Type-1, $w2 - w3$ & Type-2, $w2 - w3$ \\
$z$ & 2788 & 2788 & 3.509 $\pm$ 0.007 & 3.628 $\pm$ 0.006\\
$L$[\ion{O}{iii}]5007 & 1244 & 1244 & 3.495 $\pm$ 0.001 & 3.648 $\pm$ 0.009\\
$F$(H$\alpha$/H$\beta$) & 1616 & 1616 & 3.518 $\pm$ 0.009 & 3.633 $\pm$ 0.008\\
$m_{w4}$ & 2714 & 2714 & 3.462 $\pm$ 0.006 & 3.668 $\pm$ 0.006\\[0.2ex]
\hline\hline
\end{tabular}
\label{WISERefinedZooSamples}
\end{table*}

One would wish to explore the SN rate in these samples at $d < $ 10 kpc. However, the currently available SN samples are far too small to allow this kind of investigation. We have to rely on an alternative star-formation indicator. We can use $WISE$ $w2 - w3$ colour to measure the activity of star-formation, assuming that our objects $WISE$ colours are dominated by the same source (either AGN or the stars): the higher $w2 - w3$, the more star-formation \citep{Coziol2015}.

For the refined samples, we see a strong difference in the $w2 - w3$ colour. The $w2 - w3$ is significantly larger in Type-2 AGN compared to Type-1 AGN for all four matchings. An example is the $w4$-matched Type-1 and Type-2 refined samples, $w2 - w3$=2.962 $\pm$ 0.03 for Type-1s and $w2 - w3$=3.674 $\pm$ 0.029 for Type-2s. The other matchings give very similar results. This strongly supports the observed larger number of SN detections around Type-2 AGN.

Also if redoing the entire pairwise matching and analysis only using objects having the infrared WISE emission dominated by dust emission near stars using either $w1 - w2$ $<$ 0.8 \citep{Assef} and or $w1 - w2$ $<$ 0.5 \citep{Wright}, the conclusion stays equally true.

\subsection{Biases in morphology?}\label{sec:MorphologyBias}
The Galaxy Zoo Data Release 1 morphologies (for those that have, $\sim$ 90 \% of all galaxies) fall into three categories: Spiral, Elliptical and Uncertain. The higher redshift, the more difficulties a Galaxy Zoo volunteer has to recognize a certain morphology. Therefore, it might be difficult for a Galaxy Zoo-observer to recognize a Spiral at higher $z$, especially if there is a strong light from the nucleus. Some Spirals can fall into the category of ``Uncertain'' due to the strong light in Type-1 AGN. 

However, this bias cannot cause misclassifications in the other direction. While a Spiral can be classified as ``Uncertain'' it is very unlikely an ``Uncertain'' galaxy will be classified as a Spiral. Therefore, what is in Galaxy Zoo classified as a Spiral, is very likely to be a Spiral as voted by hundreds of Galaxy Zoo volunteers. 

Any redshift-dependent biases as the morphology-classification bias is removed by the use of redshift-matched (all-morphologies) samples. They show significant differences between Type-1 and Type-2 AGN.  We attempted doing the same analysis for redshift-matched Type-1 and Type-2 spiral hosts, but given the extremely small statistics it was not possible. We can only hope that future data releases will permit us to do this final, but very important test. This motivated us to use the alternative star-formation indicator as in Section \ref{sec:AltSFR}.

As an additional test we also visually classified the de Vaucoleurs-Buta stage in the De Vaucouleurs Revised Hubble-Sandage Classification System of the Type-1 and Type-2 AGN in the refined samples to see if any insight could be gained about the relative age of the stellar populations without seeing difference.

\subsection{Biases in SN-AGN matches?}\label{sec:SNbiasiPTF}

When matching our AGNs with the iPTF SNe, some biases may be introduced which could affect the results. Two sources of bias are considered:

\begin{itemize}
\item On 2014 June 17, automatic filtering was introduced in the iPTF SN scanning software in order to save known AGNs as so called \textit{Nuclear} objects before a human scanner could begin vetting the candidates. The AGN identification done from 2014 June and onwards was based on SDSS DR10 data. In 2015 February, QSOs from SDSS DR 12 were added (Yi Cao, personal communication).

\item The match radius used to tie changes in brightness to a certain transient is $1''$. This means that if the angular separation between two transient sources exceeds $1''$, they are considered to be different sources. This could lead to a potential skewness in SN detections near AGNs, arising from confusing AGNs with SNe if they reside $1''$ or less from each other.
\end{itemize}

We avoid the first bias, throughout this work, by only considering SNe discovered \textit{before} 2014 June 17. However, we need to perform a test for the second potential bias. The impact of the second bias can be weakened by only considering SNe appearing at angular distances from AGN $>3''$. This is generously larger than what is called for by the $1''$ matching condition.

We will herein present three different tests that probe the two biases. The results support a coherent targeting of SNe around Type-1s and Type-2s. The observed differences in SNe counts presented is thus a physical effect.

\subsubsection{Suitability of the iPTF}

Evaluating the detection efficiency and completeness of a SN survey is a complex task \citep{Taylor2014}. The untargeted iPTF SN search is concentrated on finding SNe in the nearby ($z < 0.2$) universe. The lack of completeness can be evaluated by repeating our SN matching to the our galaxy samples using the extra condition e.g. $z_{SN} < 0.1$. Applying this condition when matching gives 1 SN around 11632 AGN Type-1s (detection fraction $\sim$ 8.6 $\ast$ 10$^{-5}$) and 50 SNe around 77708 AGN Type-2s (detection fraction $\sim$ 6.4 $\ast$ 10$^{-4}$) for $d < 100$ kpc and the largest host samples. The estimated size of effect is $\sim$ 4.1$\sigma$ in this case. This indicates that the iPTF catalogue of SNe used by us is sufficiently complete for this investigation.

We remind that all 2190 SNe used in our study are spectroscopically classified (Sect. \ref{sec:iPTF}). Since not all SN candidates found by iPTF with the P48 telescope were eventually spectroscopically classified, we have to examine whether a bias against classification of SNe residing in either AGN Type 1 or Type 2 hosts was somehow introduced when SN classification targets were selected. Such bias, if present, would adversely affect the usefulness of our SN sample. 

To examine if there is a bias, we manually vetted all transients found during our 2009-2014 period in vicinity (on the plane of the sky) of our AGN sample galaxies. The search radius around each galaxy was set as the apparent size of 100 kpc at the distance of each galaxy. We noted that, around both types of AGNs, about 1/3 of the likely SN candidates found were eventually classified spectroscopically. The similarity in fraction of SN candidates spectroscopically classified in and around our AGN Type-1 and Type-2 hosts, respectively, shows that our SN sample is suitable for our study.

\subsubsection{Test with a $r > 3''$ cut}

Adding the harsh condition of at least $3''$ separation when matching our 2190 SNe gives in total 1 SN around 11632 Type-1s (detection fraction $\sim$ 8.6 $\ast$ 10$^{-5}$) and 46 SNe around 77708 Type-2s (detection fraction $\sim$ 5.9 $\ast$ 10$^{-4}$) for $d < 100$ kpc and the largest host samples. The estimated size of effect is $\sim$ 4.1$\sigma$ in this case. 

Visual inspection of iPTF discovery images of SNe found $2.0'' < r < 3.0''$ from an AGN show it is generally easy for a scanner to unambiguously tell if a SN candidate is separated from the central region of a galaxy. The demonstration that even such a conservative limit as $r > 3''$ can maintain a $\sim$ 4.1$\sigma$ effect suggests that we are seeing a physical effect and not a bias effect.

\subsubsection{Test using foreground and background objects, $d < 100$ kpc}\label{sec:largescale1}

We can also explore possible detection effects by looking at the number of foreground and background SNe. If no detection probability effects are at play, the number of background/foreground SNe should be the same near Type-1 and Type-2 AGN. We set the redshift criterion  $|\Delta z|$ $>$ 0.07 and search for SNe  within 100 kpc of projected distance on the sky around our redshift-matched samples. We find no SNe in apparent vicinity of any AGN, in the foreground or background, within projected distance $d < 10$ kpc. No significant difference in SNe counts are found around the largest host samples either.

\subsubsection{Test using large-scale environment, $100 < d < 1100$ kpc}\label{sec:largescale2}

Another way of exploring whether or not there is a different detection rate of SNe near Type-1s and Type-2s, is by counting the number of SNe at large projected separations. At sufficiently large separations one expects the AGN neither influence their surroundings nor the surroundings to influence the AGN significantly. On the other hand, some recent articles \citep[e.g.][]{Donoso2014} suggest unobscured (presumably Type-1s) AGN reside in less dense large-scale (Mpcs) environments than obscured (presumably Type-2s) AGN. Such an effect would influence the stellar mass and also the SN rate that is expected to be higher in the obscured AGN case.

We use the $|\Delta z| < $ 0.003 but now select SNe within $100 < d < 1100$ kpc, where effects are expected to disappear. For redshift-matched samples of 10146 Type-1 and Type-2 AGN, we see that for Type-1 AGN we find 22 SNe (detection fraction $\sim$ 2.2 $\ast$ 10$^{-3}$), and for Type-2 AGN we find 15 SNe (detection fraction $\sim$ 1.5 $\ast$ 10$^{-3}$), yielding no significant difference in the large-scale environment. For none of the samples we find significant differences in the detected SNe rate for Type-1s and Type-2s in an annulus of $100 < d < 1100$ kpc, supporting our claim that the detection of SNe near Type-1s and Type-2s in iPTF is not biased in favour of any of the two AGN types.

\section{Statistical analysis}\label{sec:Statistical}

A statistical analysis of the results can be done in several different ways. We can get an estimate of the size of effect without assuming anything about the underlying distributions (Section \ref{sec:unc}). A detection signal from estimating at what level the reported difference between the two samples is just by chance can be obtained by assuming the two AGN samples have equal SN rates as the null hypothesis (Section \ref{sec:nulldist}). But, ultimately hypothesis testing should be used to estimate the probability $p$ for the hypothesis to be true and tell us whether or not we can reject the null hypothesis for the nominal value $\alpha$=0.05 (Section \ref{sec:pairwise}).

\subsection{Estimate of the size of effect}\label{sec:unc}
Let the random variables $X_1\sim Po(N_1\lambda_1)$ and $X_2\sim Po(N_2\lambda_2)$ denote the number of SNe in galaxy sample no. $1$ (Type-1 AGN) and $2$ (Type-2 AGN), respectively, and let $N_i\sim Po(\mu_i)$, be the total number of galaxies observed in sample no. $i,~i=1,2$. Here, the parameter $\lambda_i$  denotes the average rate of SNe per galaxy in the Type-$i$ AGNs, while $\mu_i$ denotes the number count of Type-$i$ AGNs in the observed volume. The random variable

\begin{equation}
Z = \frac{X_1}{N_1} - \frac{X_2}{N_2},
\label{eq.z}
\end{equation}

\noindent
then describes the difference between the rates of SNe in the two samples. In order to put the observed difference in context, we define the ratio

\begin{equation}
\beta = \frac{|\mathbb{E}(Z)|}{\sqrt{Var(Z)}},
\label{eq.etos}
\end{equation}

\noindent
which is a measure of the expected size of the difference in units of the standard deviation of the distribution of $Z$. Note that, in general, $Z$ is not normally distributed, wherefore a detection at a level of, say, $3\sigma$ cannot directly be compared with a $3\sigma$ detection where the distributions of interest are normal. In our case, however, the difference should be relatively small.\\

\noindent
Since the difference in Eq. (\ref{eq.z}) involves ratios of two random variables, the computation of the expectation and variance of $Z$ is not straightforward and we will use Gauss's approximation formulae (first order) to derive an explicit expression of Eq. (\ref{eq.etos}). We have that 

\begin{eqnarray}
Var(Z) &=& Var\left(\frac{X_1}{N_1} - \frac{X_2}{N_2}\right) = \nonumber \\
& = & Var\left(\frac{X_1}{N_1} \right) + Var\left(\frac{X_2}{N_2} \right),
\end{eqnarray}

\noindent
where it is assumed that $X_1/N_1$ and $X_2/N_2$ are independent. Now, 

\begin{eqnarray}
Var\left(\frac{X_i}{N_i} \right) &\approx& \frac{1}{\mu_i^2}\cdot Var(X_i) + \nonumber \\
&+& \frac{\mathbb{E}(X_i)^2}{\mu_i^4}\cdot Var(N_i) - \nonumber \\
&-& 2\cdot \frac{1}{\mu_i} \frac{\mathbb{E}(X_i)}{\mu_i^2}\cdot C(X_i,N_i).
\end{eqnarray}

\noindent
By the fact that $X_i|N_i=n \sim Bin(n,\lambda_i)$, the law of total expectation gives that $\mathbb{E}(X_i) = \mathbb{E}(\mathbb{E}(X_i|N_i)) = \mathbb{E}(N_i\lambda_i) = \mu_i\lambda_i$. Similarly, for the variance we have that 

\begin{eqnarray}
Var\left(X_i \right) &=& \mathbb{E}(Var(X_i|N_i)) + Var(\mathbb{E}(X_i|N_i)) = \nonumber \\
&=& \mathbb{E}(N_i\lambda_i(1-\lambda_i)) + Var(N_i\lambda_i) = \nonumber \\ 
&=& \mu_i\lambda_i(1-\lambda_i) + \mu_i\lambda_i^2 = \mu_i\lambda_i,
\end{eqnarray}

\noindent
while the covariance also amounts to $C(X_i,N_i) = \mu_i\lambda_i$. Hence, we have that 

\begin{eqnarray}
Var\left(\frac{X_i}{N_i} \right) \approx \frac{\lambda_i}{\mu_i} + \frac{\lambda_i^2}{\mu_i} - 2\cdot \frac{\lambda_i^2}{\mu_i} = \frac{\lambda_i(1-\lambda_i)}{\mu_i}.
\end{eqnarray}

\noindent
For the expectation of $Z$, we have that $\mathbb{E}(Z) = \mathbb{E}(X_1/N_1-X_2/N_2) \approx \lambda_1-\lambda_2$. It is noted that also for the second order approximation, $\mathbb{E}(Z) \approx \lambda_1-\lambda_2$  Thus, we have 

\begin{eqnarray}
\beta &=& \frac{|\mathbb{E}(Z)|}{\sqrt{Var(Z)}} \approx \frac{|\lambda_1-\lambda_2|}{\sqrt{\frac{\lambda_1(1-\lambda_1)}{\mu_1} + \frac{\lambda_2(1-\lambda_2)}{\mu_2}}} =\nonumber \\
&=& \frac{\sqrt{\mu_1\mu_2}|\lambda_1-\lambda_2|}{\sqrt{\lambda_1(1-\lambda_1)\mu_2+ \lambda_2(1-\lambda_2)\mu_1}}.
\label{eq.ratio}
\end{eqnarray}

\noindent
Now, a plug-in estimate of the ratio in Eq. (\ref{eq.ratio}) is given by the expression

\begin{equation}
\widehat{\beta} \approx \frac{\sqrt{\widehat{\mu}_1\widehat{\mu}_2}|\widehat{\lambda}_1-\widehat{\lambda}_2|}{\sqrt{\widehat{\lambda}_1(1-\widehat{\lambda}_1)\widehat{\mu}_2+ \widehat{\lambda}_2(1-\widehat{\lambda}_2)\widehat{\mu}_1}},
\label{eq.rest}
\end{equation}

\noindent
where $\widehat{\lambda}_i, \widehat{\mu}_i,~i=1,2$ are estimated from the observed numbers. For example, for the full samples and $d<100$ kpc, we have that  $\widehat{\mu}_1=11632,~ \widehat{\mu}_2=77708,~\widehat{\lambda}_1=1/11632$, and $\widehat{\lambda}_2=59/77708$ (see Table \ref{Largest}). Thus, we obtain 

\begin{equation}
\widehat{\beta} \approx 5.1,
\label{eq.example}
\end{equation}

\noindent
i.e., the observed difference is more than $5\sigma$ away from $\lambda_1-\lambda_2=0$, as defined by Eq. (\ref{eq.etos}).\\

\noindent
Since the SN rates generally are very small, i.e., $\lambda \sim10^{-4}-10^{-3}$, the factor $1-\lambda_i \simeq 1$ in the expression for $Var(Z)$ and we may take $N_1=n_1$ and $N_2=n_2$ as fixed. Then, $Z$ reduces to 

\begin{equation}
Z=\frac{X_1}{n_1}-\frac{X_2}{n_2} 
\end{equation}

\noindent
and

\begin{eqnarray}
\widehat{\beta} &=& \widehat{\frac{|\mathbb{E}(Z)|}{\sqrt{\widehat{Var(Z)}}}} = \frac{\sqrt{n_1n_2}|\widehat{\lambda}_1-\widehat{\lambda}_2|}{\sqrt{\widehat{\lambda}_1n_2+ \widehat{\lambda}_2n_1}} = \nonumber \\
&=&\frac{|x_1/n_1-x_2/n_2|}{\sqrt{x_1/n_1^2+x_2/n_2^2}},
\label{eq.red}
\end{eqnarray}

\noindent
where $x_i$ is the total number of SNe observed in the Type-$i$ AGNs. For $x_1=1,~x_2=59,~n_1=11632$, and $n_2=77708$, we have $\widehat{\beta}\approx 5.1$, as the result in \mbox{Eq. (\ref{eq.example}).} For the hypothesis testing discussed in Sect. 1.7.3, we make the assumption that $n_1$ and $n_2$ are fixed.\\

\subsection{Statistic based on the null distribution}\label{sec:nulldist}
\noindent
Let's hypothesise that the SN rates of the two AGN samples are equal. At what level will the observed difference then just be due to chance? By taking the null hypothesis to be\\ 

\noindent
$\hphantom{00000000}H_0:~\lambda_1=\lambda_2~(=\lambda_0)$ versus\\ 
$\hphantom{00000000}H_1:~\lambda_1 \neq \lambda_2$,\\

\noindent
where $\lambda_0$ is the common SN rate, we have instead that $X_1\sim Po(N_1\lambda_0)$ and $X_2\sim Po(N_2\lambda_0)$. Consequently, the variance of the random variable $Z=X_1/N_1-X_2/N_2$ becomes 

\begin{equation}
Var(Z) \approx \lambda_0(1-\lambda_0) \left(\frac{1}{\mu_1} + \frac{1}{\mu_2} \right).
\end{equation}

\noindent
The expression of the ratio $\beta$, defined as $\beta=|\lambda_{\mathrm{diff}}|/\sqrt{Var(Z)}$ where $\lambda_{\mathrm{diff}}$ denotes the observed difference $\widehat{\lambda}_1-\widehat{\lambda}_2$, then equals

\begin{equation}
\beta = \frac{|\lambda_{\mathrm{diff}}|}{\sqrt{Var(Z)}} \approx \frac{|\lambda_{\mathrm{diff}}|}{\sqrt{\lambda_0(1-\lambda_0)\cdot \left(\frac{1}{\mu_1} + \frac{1}{\mu_2} \right)}}.
\end{equation}

\noindent
Following the steps is Sect. \ref{sec:unc}, we have that 

\begin{equation}
\widehat{\beta} = \frac{|\lambda_{\mathrm{diff}}|}{\sqrt{\widehat{Var(Z)}}} \approx \frac{|\widehat{\lambda}_1-\widehat{\lambda}_2|}{\sqrt{\widehat{\lambda}_0(1-\widehat{\lambda}_0)\cdot \left(\frac{1}{\widehat{\mu}_1} + \frac{1}{\widehat{\mu}_2} \right)}},
\end{equation}

\noindent
where an estimate of $\lambda_0$ is given by

\begin{equation}
\widehat{\lambda}_0 = \frac{x_1+x_2}{n_1+n_2}.
\end{equation}

\noindent
Hence, for the full, unmatched samples ($d<100$ kpc) we have that $\widehat{\lambda}_0 = (1+59)/(11632+77708) \simeq 6.72\times 10^{-4}$ and $\widehat{\beta} \approx 2.6$. This is a more conservative measure of detection, i.e. a difference in the SN rates is detected close to the $2.6\sigma$-level.\\

\noindent
Finally, we have that 

\begin{eqnarray}
\widehat{\beta} &=& \frac{|\widehat{\lambda}_1-\widehat{\lambda}_2|}{\sqrt{\widehat{\lambda}_0/n_1+ \widehat{\lambda}_0/n_2}} = \nonumber \\
&=& \sqrt{\frac{n_1n_2}{x_1+x_2}} \cdot |x_1/n_1-x_2/n_2|,
\label{eq.red2}
\end{eqnarray} 

\noindent
under the assumption that $Z=X_1/n_1 - X_2/n_2$. It is noted that for this approximation, the estimates of $\beta$ as given by Eq. (\ref{eq.red}) and Eq. (\ref{eq.red2}) are equal for $n_1=n_2$.\\

We report this estimate of the signal of detection alongside with the size of effect and p-value in the captions to Table \ref{SupernovaAll}.

\subsection{Statistical hypothesis testing of the SN counts in pairwise matched samples}\label{sec:pairwise}
The number of SNe in the host galaxies of AGN should, within the realm of the Unification model, not depend on the type of AGN. We have performed a statistical test of this hypothesis. Recall that the SN rate (read proportion/success probability) $\lambda$ in a host galaxy population of size $n$ should be binomially distributed. However, since $\lambda < 0.003 \ll 0.1$ in all observed cases, $Bin(n,\lambda)\approx Po(n\lambda)$ to a very high degree. Therefore,  we assume that the number of SNe believed to be associated with the Type-1 and Type-2 AGN samples are, respectively, observations of the random variables $X_1\sim Po(n_1\lambda_1)$ and $X_2\sim Po(n_2\lambda_2)$. Here,  $n_i,~i=1,2$ is the size of the Type-$i$ AGN sample and $\lambda_i$ is the corresponding rate of SNe in the Type-$i$ AGN population, uniformly corrected for the biases discussed above.  Furthermore, it is assumed that $X_1$ and $X_2$ are independent. Let the null hypothesis be\\

\noindent
 \hphantom{00000000}$H_0: \lambda_1 = \lambda_2$, against the alternative\\
 \hphantom{00000000}$H_1: \lambda_1 \neq \lambda_2$.\\
 
\noindent 
The exact conditional test (C-test) for comparing two Poisson means by Przyborowski \& Wilenski (1940) is known to be overly conservative, i.e., the chance of failing to reject a false null hypothesis is higher than the nominal level. We have therefore chosen to perform a test based on estimated p-values instead (Krishnamoorthy \& Thomson, 2004). The pivot statistics for $\lambda_1-\lambda_2=0$ is given by

\begin{equation}
T_{X_1,X_2} = \frac{X_1/n_1-X_2/n_2}{\sqrt{\hat{V}_{X_1,X_2}}},
\end{equation}

\noindent
where 

\begin{equation}
\hat{V}_{X_1,X_2} = \frac{X_1}{n_1^2} + \frac{X_2}{n_2^2}
\end{equation}

\noindent
is the unbiased variance estimator of the standardised difference $X_1/n_1 - X_2/n_2$. 
The p-value for the two-sided test is then given by 

\begin{equation}
p = P(|T_{X_1,X_2}|\ge |T_{x_1,x_2}|~| H_0),
\end{equation}

\noindent
where $T_{x_1,x_2} = (x_1/n_1-x_2/n_2)/\sqrt{\hat{V}_{x_1,x_2}}$ is the observed value of $T_{X_1,X_2}$ and $x_1$ and $x_2$ are the observed numbers of SNe in the Type-1 and Type-2 sample, respectively. Hence, the p-value is estimated by the expression (see Krishnamoorthy \& Thomson, 2004)

\begin{equation}
p=\sum\limits_{k_1=0}^{\infty} \sum\limits_{k_2=0}^{\infty} \frac{e^{-n_1\hat{\lambda}_2}(n_1\hat{\lambda}_2)^{k_1}}{k_1!} \frac{e^{-n_2\hat{\lambda}_2}(n_2\hat{\lambda}_2)^{k_2}}{k_2!}I_{|T_{k_1,k_2}| \ge |T_{x_1,x_2}|},  
\end{equation}

\noindent
where $I_{a \ge b}$ denotes the indicator function such that

\begin{equation}
I_{a \ge b} = \left\{
\begin{array}{lll}
1,&& a \ge b,\\
0,&& a < b,
\end{array}\right.
\end{equation}

\noindent
and $\hat{\lambda}_2=(x_1+x_2)/(n_1+n_2)$ is the estimate of $\lambda_2$ (see eq. 17). Note that under the null hypothesis, $\hat{\lambda}_1 = \hat{\lambda}_2$. Also, we have that $T_{0,0} = 0$. The null hypothesis is then rejected at the significance level $\alpha$ if $p\le \alpha$. We use the nominal value $\alpha=0.05$.

We results from the hypothesis test are reported in the Table \ref{SupernovaAll}.

\section{Physical implications on AGN obscuration}\label{sec:implications}

It might be interesting to speculate about the origin of the obscuration. It seems fairly natural to assume that a larger number of massive, young stars in the galaxy (dying as core-collapse SNe) leads to a larger production of dust. The first suspicion of that galactic-scale dust is responsible for some of the obscuration in Type-2 AGN came from the discovery that Type-1 AGN rarely are found in edge-on systems \citep{Keel1980}. Later, it was found that Seyfert-2 have more dust lanes and dust patches near their nuclei than do Seyfert-1 AGN \citep{Malkan1998}, in contradiction with the Simplest Unification where all obscuration is caused by the torus -- the doughnut-like dust structure that surrounds the AGN engine, no larger than a few hundred parsecs. The column density of galactic dust outside the most central 500 pc was estimated to $N_{H}$ $\sim  4 \ast 10^{22}$ cm$^{-2}$, close to the frequently used limit in column density to separate between obscured and unobscured AGN (unobscured $N_{H} < 10^{22}$ cm$^{-2}$). 

\begin{figure*}
 \centering
   \includegraphics[scale=.5]{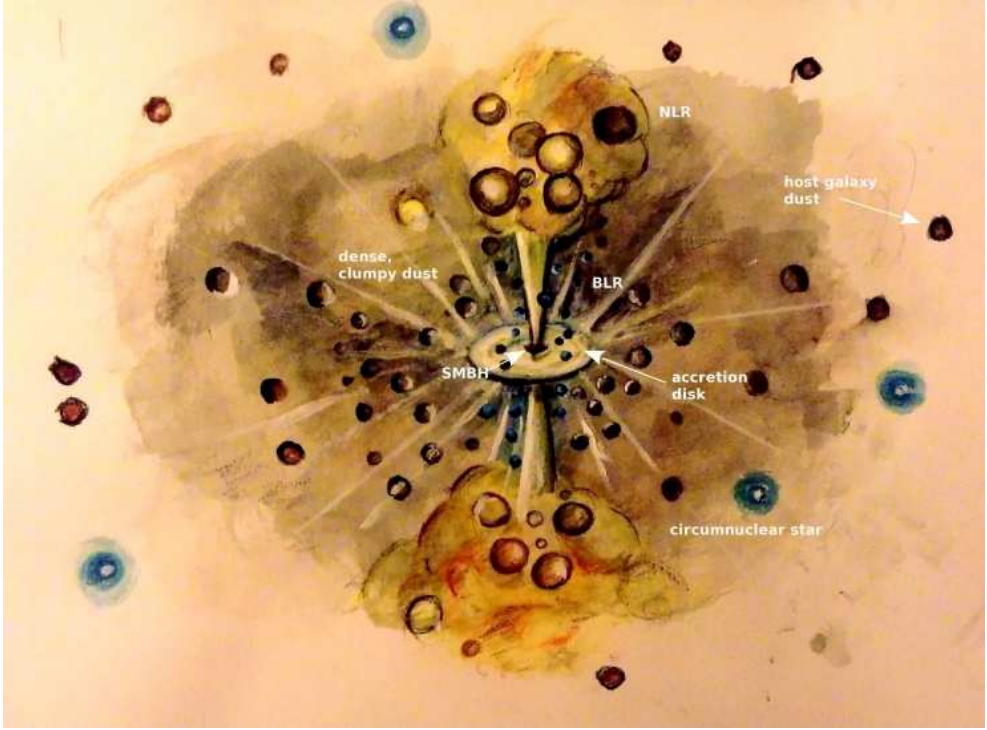}
     \caption{A cartoon of the AGN Unification model. The line-of-sight towards the BLR and the accretion disk determines whether the observer sees a Type-1 or Type-2 AGN. Any sufficiently dense dust can obscure the Type-1 AGN, so that only a Type-2 (or a partially obscured AGN) is observed. The cartoon shows all components needed to describe the AGN and connection between the two classes: the SMBH, the accretion disk, the BLR, the NLR, obscuring dust and eventual jets. Near the central engine, at scales of a few parsecs, the dust is clumpy and dense, while further out, on scales of a few hundreds of parsecs and beyond, the dust is considerably less dense, but can still cause obscuration of the central engine. While the luminosity of the engine can influence the parsec-size dust-sublimation radius according to \cite{Buchner2016}, the star-formation history of the host galaxy influences the large-scale obscuration.}
     \label{Unification}
     \end{figure*}

Recent works support the idea that the host galaxy has enough dust \citep{Burtscher2016} to cause obscuration of the broad-line region. DiPompeo et al. (2016) estimate the fraction of IR-selected ``obscured'' AGN that are obscured by dust outside the torus to $\sim$ 25\%, while Buchner et al. (2016) estimate that 40 \% of all AGN have considerable host galaxy obscuration. But objects with large measured column densities $N_{H} > 10^{23.5}$ cm$^{-2}$ cannot be explained with host galaxy obscuration and need heavy obscuration around the nucleus. We include a cartoon of an AGN, Figure \ref{Unification}, that shows all components needed to describe the AGN Unification.

If we compare the $L$[\ion{O}{iii}]5007-matched Type-1 galaxies ($N$=6252) and Type-2 galaxies ($N$=6252) used for SN counts, we can get a rough estimate of the fraction of AGN that are obscured only by the torus using the na\"ive assumption that there is a one-to-one correspondence in expected properties between the objects within each $L$[\ion{O}{iii}]5007-matched pair of host galaxies, i.e. same Galaxy Zoo morphology, and with measured exponential fit scale radius \& $M_{w4}$ within 20\% error. The fraction of objects fulfilling these criteria shows that 10\% can be explained purely by (parsec-size) torus obscuration and agree with the predictions from the Simplest Unification about identical host galaxy properties. (Using the $L$[\ion{O}{iii}]5007-matched refined samples based on spiral face-on hosts, the corresponding na\"ive fraction is 25\%.) This also means that for majority of Type-2 AGN, up to 90\%, the obscuration must stem from larger scales: host galaxy obscuration and/or large-scale environment in which the host galaxies reside.

The finding that $M_{w4}$-matched Type-1 AGN are significantly more luminous in [\ion{O}{iii}]5007 than the Type-2 AGN in host-matched samples, asks for a physical connection between the AGN classes that goes beyond pure obscuration. Perhaps the two important differences between Type-1 and Type-2 AGN that we find on a population level -- the different average AGN luminosity and the different stellar ages of the host galaxies -- can be explained within the framework of an evolutionary scenario \citep{Sanders1988,Hopkins2006} where the Type-2 AGN are those with obscured by heavy star formation during the initial AGN phase after the merger, where the AGN becomes stronger and stronger until the ``blow-out phase'' is reached, later leading to a strong, naked ``unobscured'' AGN with less dust and gas. The AGN environment also supports this scenario through differences in neighbour counts \citep{Dultzin,Krongold2002} and neighbour properties \citep{Koulouridis2013,VillarroelKorn2014}.

The torus could be a heavily star-forming molecular disk at $r < 100$ pc where the SNe keep up the geometrical thickness of the dust disk \citep{Wada2002,Wada2016} and is more likely to be heavily star-forming in gas-rich galaxies with younger stellar populations. The presence of starbursts \citep{Davies2006} and past starbursts \citep{Davies2014} in AGN in the most central regions support the view of heavy star formation in the most central 100 pc. But the observed difference in luminosity between the $M_{w4}$-matched Seyfert-1 and Seyfert-2 galaxies perhaps means we are dealing with a receding molecular disk. Continued theoretical efforts into modeling the torus are therefore needed.

\section{Conclusions}\label{sec:conclusions}

Using SNe from the iPTF and galaxies from the SDSS, we have searched for possible physical differences between Type-1 and Type-2 AGN. Simultaneously, we have also carefully matched Type-1 and Type-2 AGN host galaxies to gain information on the AGN engine.

We find that:

\begin{enumerate}

\item The two AGN classes differ in term of SN counts in their hosts. Type-2 AGN hosts have a larger number of SNe. This differences appears to originate in a difference between stellar ages (and not only stellar masses) and more recent star formation in Type-2 AGN hosts. Star-forming galaxies have more recent star formation than Type-2 AGN host.

\item Based on equal [\ion{O}{iii}]5007 line widths in Type-1 and Type-2 AGN (significantly wider than in star-forming hosts), there are good reasons to accept an isotropy in the [\ion{O}{iii}]5007 distribution as long assumed by the Simplest Unification.

\item Type-1 and Type-2 AGN hosts that are dominated by star formation $m_{W1} - m_{W2} < 0.5$ and are matched in cold dust emission have strong differences in NLR luminosity $L$[\ion{O}{iii}]5007 -- Type-1 AGN being 10 times more luminous.

\end{enumerate}

Summarizing, we identify two more factors beyond the viewing angle -- AGN luminosity and the age of stellar populations -- making Unification not only a question of obscuration but also one of generation. Finally, we conclude the great potential of iPTF and surveys like the upcoming Zwicky Transient Facility (ZTF), for resolving the question with larger SN samples. The ZTF \citep{Smith2014} will use the same telescope as the iPTF, but with a larger field of view, and is expected to start in 2017.



\begin{table*}[t]
\caption{SN counts around different host galaxies. We classify the SNe into three different types: SNe Type Ia (marked as 'SNIa'), core-collapse ('c-c') and 'unknown', and mark the numbers of SNe of each type in parenthesis at the form 'N (SNIa$|$c-c$|$u)' at each line. The detection fraction $f$ = $N_{SN}/N_{gal}$ indicated for each sample as the ratio between number of observed SNe and the given galaxy sample size. A strong difference in SN counts between Type-1 and Type-2 AGN at d $<$ 100 kpc is apparent. A two-sample hypothesis test \citep{Krishnamoorthy} at significance level $\alpha= 0.05$ gives the p-value $\sim$ 2.1 $\ast 10^{-5}$. The size of effect is estimated to be $\sim$ 5.1$\sigma$. The signal of detection $\sim$ 2.6$\sigma$ if using a statistics based on null distribution (Sect. \ref{sec:nulldist}). The difference is significant for both maximum projected distances (10 and 100 kpc). In the numbers of SNe around spiral-host Type-2 AGN and star-forming objects a significant size of effect ($\sim$ 3.3$\sigma$), or a 2.8 $\sigma$ signal of detection, can be found at d $<$ 10 kpc in the number of core-collapse SNe. Using a hypothesis test \citep{Krishnamoorthy} where the null hypothesis H$_{0}$ is that two samples have the same SN counts, we estimate significance levels of statistical differences for the spiral hosts AGN. We reject H$_{0}$ at the nominal value $\alpha$=0.05 and conclude that AGN hosts have fewer SNe than star-forming galaxies at d $<$ 10 and 100 kpc. Comparing spiral-host Type-1 and Type-2 AGN we get $p$=0.06, border-line significant but not enough to reject H$_{0}$, showing the need of larger samples.}
\centering
{\tiny
\begin{tabular}{|c c c c c c c|}
\hline\hline
\multicolumn{7}{c}{Total SN counts.} \\
\hline\hline
Max distance & Type-1 AGN (11632) & $f$ & Type-2 AGN (77708) & $f$ & Star-forming (137489) & $f$\\
d $<$ 10 kpc & 0 & 0 & 39 $(21|16|2)$ & 5 $\ast 10^{-4}$ & 117 $(58|56|3)$ & 8.5 $\ast 10^{-4}$\\
d $<$ 100 kpc & 1 $(1|0|0)$ & 8.6 $\ast 10^{-5}$ & 59 $(36|20|3)$ & 7.6 $\ast 10^{-4}$ & 152 $(73|76|3)$ & 1.1 $\ast 10^{-3}$\\[0.2ex]
\hline\hline
\multicolumn{7}{c}{SN counts around spiral hosts.} \\
\hline\hline
Max distance & Type-1 AGN (1864) & $f$ & Type-2 AGN (36720) & $f$ & Star-forming (49072) & $f$ \\
d $<$ 10 kpc & 0 & 0 & 25 $(12|11|2)$ & 6.8 $\ast 10^{-4}$ & 59 $(24|34|1)$ & 1.2 $\ast 10^{-3}$\\
d $<$ 100 kpc & 0 & 0 & 39 $(22|14|3)$ & 1.1 $\ast 10^{-3}$ & 77 $(33|43|1)$ & 1.6 $\ast 10^{-3}$\\[0.2ex]
\hline\hline
\end{tabular}}
\label{Largest}
\end{table*}

\newpage

\begin{table*}[t]
\caption{SN types and counts. We compare SNe around host galaxies within two different distances, 10 kpc and 100 kpc. We do it (i) for mixed Hubble types, (ii) for only spiral-hosts, and in four types of matchings. For Type-1 and Type-2 AGN in spiral hosts the number of objects are too few. We classify the SNe into three different types: SNe Type Ia (marked as 'SNIa'), core-collapse ('c-c') and 'unknown', and mark the numbers of SNe of each type in parenthesis at the form 'N (SNIa$|$c-c$|$u)' at each line. We assume the null hypothesis H$_{0}$ that the total SN rates are the same for each pair of matched samples against the alternative hypothesis $H_{1}: \lambda_1\neq \lambda_2$ and perform a two-sample test \citep{Krishnamoorthy} at significance level $\alpha=0.05$, see Sect. \ref{sec:pairwise}.}
\centering
{\tiny
\setlength\tabcolsep{2pt}
\begin{tabular}{|c || c | c c |c || c | c c |c|}
\hline\hline
\multicolumn{9}{c}{SNe within 10 kpc.} \\
\hline\hline
All hosts & N of galaxies / sample & Type-1 AGN* & Type-2 AGN &  H$_{0}$ rejected ($p\le0.05$) & N of galaxies / sample & Type-2 AGN* & Star-forming & H$_{0}$ rejected ($p\le0.05$)\\
$z$ & 10146 & 0 & 4 $(2|2|0)$ & yes ($p=4.1 \times 10^{-2} $) & - & - & - & - \\  
$L$[\ion{O}{iii}]5007 & 6252 & 0 & 0 & no ($p=1$) & - & - & - & - \\  
$M_{w4}$ & 8506 & 0 & 6 $(6|0|0)$ & yes ($p=9.6 \times 10^{-3}$) & - & - & - & - \\  
$r_{exp}$ & 7218 & 0 & 2 $(2|0|0)$ & no ($p=2.0 \times 10^{-1}$) & - & - & - & - \\  
\hline
Spiral hosts & N of galaxies / sample & Type-1 AGN* & Type-2 AGN &  H$_{0}$ rejected ($p\le0.05$) & N of galaxies /sample & Type-2 AGN* & Star-forming & H$_{0}$ rejected ($p\le0.05$)\\
$z$ & - & - & - & - & 13067 & 13 $(6|7|0)$ & 25 $(13|12|0)$ & no ($p=5.2 \times 10^{-2}$) \\  
$L$[\ion{O}{iii}]5007 & - & - & - & - & 7705 & 5 $(3|2|0)$ & 9 $(5|3|1)$ & no ($p=3.0 \times 10^{-1}$) \\
$M_{w4}$ & - & - & - & - & 13108 & 2 $(2|0|0)$ & 9 $(4|5|0)$ & yes ($p= 4.0 \times 10^{-2} $) \\  
$r_{exp}$ & - & - & - & - & 10827 & 6 $(3|3|0)$ & 14 $(5|9|0)$ & no ($p=7.9 \times 10^{-2}$) \\  
\hline\hline
\multicolumn{9}{c}{SNe within 100 kpc.} \\
\hline\hline
All hosts & N of galaxies / sample & Type-1 AGN* & Type-2 AGN &  H$_{0}$ rejected ($p\le0.05$) & N of galaxies / sample & Type-2 AGN* & Star-forming & H$_{0}$ rejected ($p\le0.05$)\\
$z$ & 10146 & 1 $(1|0|0)$ & 5 $(3|2|0)$ & no ($p=1.3 \times 10^{-1}$) & - & - & - & - \\  
$L$[\ion{O}{iii}]5007 & 6252 & 1 $(1|0|0)$ & 1 $(1|0|0)$ & no ($p=1$) & - & - & - & - \\  
$M_{w4}$ & 8506 & 1 $(1|0|0)$ & 9 $(8|1|0)$ & yes ($p=8.9 \times 10^{-3}$) & - & - & - & - \\  
$r_{exp}$ & 7218 & 1 $(1|0|0)$ & 3 $(3|0|0)$ & no ($p=4.4 \times 10^{-1}$)  & - & - & - & - \\  
\hline
Spiral hosts & N of galaxies / sample & Type-1 AGN* & Type-2 AGN &  H$_{0}$ rejected ($p\le0.05$) & N of galaxies / sample & Type-2 AGN* & Star-forming & H$_{0}$ rejected ($p\le0.05$)\\
$z$ & - & - & - & - & 13067 & 18 $(10|8|0)$ & 28 $(13|15|0)$ & no ($p=1.4\times 10^{-1}$) \\  
$L$[\ion{O}{iii}]5007 & - & - & - & - & 7705 & 9 $(5|3|1)$ & 10 $(5|4|1)$ & no ($p=8.4 \times 10^{-1}$) \\  
$M_{w4}$ & - & - & - & - & 13108 & 8 $(7|1|0)$ & 12 $(7|5|0)$ & no ($p=3.9 \times 10^{-1}$) \\  
$r_{exp}$ & - & - & - & - & 10827 & 11 $(6|4|1)$ & 15 $(6|9|0)$ & no ($p=4.4 \times 10^{-1}$) \\  
\hline
\hline
\end{tabular}
}
\label{SupernovaAll}
\end{table*}

\newpage

\newpage

\section{Acknowledgments}

B.V. wishes to thank Kjell Lundgren for early discussions about the work. She also wishes to thank Martin Gaskell and Robert R.J. Antonucci and  for many constructive and helpful comments, and finally acknowledge A. Magnard, C. Franck, G. von Rivia and J. Pianist for inspiring discussions. A.N. wishes to thank Ariel Goobar for useful suggestions. B.V. was funded and supported by the Center of Interdisciplinary Mathematics (Uppsala Universitet) and Erik and M\"arta Holmbergs donation from the Kungliga Fysiografiska S\"allskapet. Supernova research at the Oskar Klein Centre is supported by the Swedish Research Council and by the Knut and Alice Wallenberg Foundation. The intermediate Palomar Transient Factory project is a scientific collaboration among the California Institute of Technology, Los Alamos National Laboratory, the University of Wisconsin, Milwaukee, the Oskar Klein Centre, the Weizmann Institute of Science, the TANGO Program of the University System of Taiwan, and the Kavli Institute for the Physics and Mathematics of the Universe.
This publication makes use of data products from the Wide-field Infrared Survey Explorer, which is a joint project of the University of California, Los Angeles, and the Jet Propulsion Laboratory/California Institute of Technology, funded by the National Aeronautics and Space Administration. This research also heavily relies on the Sloan Digital Sky Survey (SDSS). Funding for SDSS-II has been provided by the Alfred P. Sloan Foundation, the Participating Institutions, the National Science Foundation, the U.S. Department of Energy, the Japanese Monbukagakusho, and the Max Planck Society.



\end{document}